\documentclass[%
reprint,
 amsmath,amssymb,
 aps, physrev,
 prl
]{revtex4-2}

\usepackage{graphicx}
\usepackage{dcolumn}
\usepackage{bm}
\usepackage{braket}
\usepackage{color}
\usepackage[normalem]{ulem}
\usepackage[bookmarks=false]{hyperref}
  \hypersetup{pdfpagemode={UseOutlines},
  bookmarksopen=true, 
  bookmarksopenlevel=0, 
  hypertexnames=false,
  colorlinks=true,
  citecolor=blue,
  linkcolor=blue,
  urlcolor=blue,
  allcolors=blue,
  pdfstartview={FitV},
  unicode,
  breaklinks=true,
  }
 
\begin{document}

\preprint{APS/123-QED}

\title{
  Strong Evidence for Three-$\bm \alpha$ Clustering in the Ground State of
  $^{\bf 12}{\bf C}$}

\author{Kazuki Yoshida}
\email{yoshidak@rcnp.osaka-u.ac.jp}
\affiliation{Research Center for Nuclear Physics (RCNP), The University of Osaka, Ibaraki 567-0047, Japan}
\affiliation{RIKEN Center for Interdisciplinary Theoretical and Mathematical Sciences (iTHEMS), RIKEN, Wako 351-0198, Japan}

\author{Masaaki Kimura}%
\email{masaaki.kimura@ribf.riken.jp}
\affiliation{RIKEN Nishina Center, Wako, Saitama 351-0198, Japan}

\date{\today}
\begin{abstract}
  The ground state of $^{12}\mathrm{C}$ has often been approximated by a mean-field picture. This conventional view has been challenged by recent nuclear theories suggesting non-negligible $\alpha$-cluster formation, but experimental evidence remains inconclusive.
  Here, we show that existing $^{12}\mathrm{C}(p,p\alpha)^{8}\mathrm{Be}$ data provide direct evidence for a pronounced $\alpha$ cluster formation in the ground state of $^{12}\mathrm{C}$.
  We analyze the data with distorted-wave impulse approximation using $\alpha$ preformation amplitudes from an unrestricted $3\alpha$ cluster model and harmonic-oscillator-based models.
  The results show that the former reproduces the measured cross sections, whereas the latter underestimate them by more than an order of magnitude. Thus, contrary to conventional expectations, the data support a nearly fully developed three-$\alpha$ cluster structure in the ground state of $^{12}\mathrm{C}$.
\end{abstract}

\maketitle

The nucleus $^{12}\mathrm{C}$ has been a representative system for studying
the cluster correlations in finite nuclei. The pronounced $\alpha$ clustering of excited states, in particular the Hoyle state, has been extensively investigated~\cite{Fujiwara1980, KanadaEnyo1998, Neff2004, Tohsaki2001,Epelbaum2012,Freer2014,Funaki2015}.

In contrast, the ground state is often taken for granted to be reasonably approximated by a $p$-shell system within a mean-field picture~\cite{Cohen1965,Kurath1973,Millener1975}.
This interpretation has been partially supported by experimental data such as single-particle spectroscopic factors of nucleon knockout reactions~\cite{Lapikas2000,Noro2020}.
However, such data do not uniquely determine the underlying many-nucleon correlations, leaving the degree of $\alpha$ cluster correlations unclear.

Recent studies are challenging this conventional picture.
Algebraic cluster models have explained the low-lying spectrum of $^{12}\mathrm{C}$ by assuming equilateral-triangular three-$\alpha$ configurations~\cite{Marin-Lambarri2014,Bijker2020}.
Modern ab-initio calculations have indicated non-negligible $\alpha$-cluster
formation in the ground state~\cite{Otsuka2022,Shen2023}, while similar indications have also been obtained in density-functional studies~\cite{Ebran2012,Marevic2019}.

To test these predictions, signatures of clustering have been inferred from
intrinsic density distributions reconstructed from electron-scattering
data~\cite{Kimura2024} and ground-state density profiles probed by proton
elastic scattering~\cite{Horiuchi2023}. In addition,
$(\alpha,\alpha')$ scattering provides a probe of the relation
between the ground state and the Hoyle state through their transition
properties~\cite{John2003,KanadaEnyo2019,Vitturi2020,Casal2021}.
However, $\alpha$ clustering is intrinsically a four-body correlation and
cannot be fully characterized by one-body densities. Moreover, transition
densities depend on both the initial and final states and therefore do not
isolate the $\alpha$ correlation in the ground state. Thus, a more direct
probe of the $\alpha$ correlation in the ground state is desirable.

The $(p,p\alpha)$ reaction provides such a probe through the cluster preformation amplitude,
which is a many-body overlap that directly measures the $\alpha+{}^{8}\mathrm{Be}$ component of the ground-state wave function. It has been applied to $^{10}$Be, $^{20}$Ne, $^{48}$Ti, and Sn isotopes~\cite{Yoshida19,Li2023,Taniguchi_48Ti,Tanaka2021}.
For $^{12}\mathrm{C}$, Mabiala et al.~\cite{Mabiala09} analyzed the $^{12}\mathrm{C}(p,p\alpha)^{8}\mathrm{Be}$ reaction using a phenomenological $\alpha$ preformation amplitude with fitted normalization, and concluded that the extracted $\alpha$ spectroscopic factor was consistent
with a conventional $p$-shell estimate.
Here, we instead derive the $\alpha$ preformation amplitude microscopically
and show that the same data require an almost fully developed three-$\alpha$
configuration rather than the conventional $p$-shell configurations.

At sufficiently high incident energies, the $(p,p\alpha)$ reaction is described within the distorted wave impulse approximation (DWIA)~\cite{Chant77,Chant83,Wakasa17}.
We adopt the factorization approximation and treat the $p$-$\alpha$ scattering as the elementary process.
Under these assumptions, the triple-differential cross section (TDX) for $\alpha$ knockout from an $L=0$ orbit leaving the ground state of $^{8}\mathrm{Be}$ is given by
\begin{align}
  \label{eq:tdx}
  \frac{d^3\sigma}{dE_p^{\mathrm{L}} d\Omega_p^{\mathrm{L}} d\Omega_\alpha^{\mathrm{L}}}
   & =
  \mathcal{J}_{LG} F_{\mathrm{kin}}^{\mathrm{L}} \frac{(2\pi)^4}{\hbar v_{p_0}}
  \left(
  \frac{2\pi \hbar^2}{\mathcal{M}_{p\alpha}^{\mathrm{t}}}
  \right)
  \frac{d\sigma_{p\alpha}}{d\Omega_{p\alpha}}
  \left|\bar{T}\right|^2,
\end{align}
where the superscript $\mathrm{L}$ indicates quantities in the laboratory frame. $\mathcal{J}_{\mathrm{LG}}$ is the Jacobian from the laboratory frame to the center-of-mass frame, and $F_{\mathrm{kin}}^{\mathrm{L}}$ is the kinematic phase-volume factor. $v_{p_0}$ is the Lorentz-invariant relative velocity between the incident proton and the target, and $\mathcal{M}_{p\alpha}^{\mathrm{t}}$ is the reduced energy of the $p$-$\alpha$ two-body system in the $p$-$\alpha$ center-of-mass frame.

The differential cross section ${d\sigma_{p\alpha}}/{d\Omega_{p\alpha}}$ for the elementary $p$-$\alpha$ scattering was calculated using the latest parametrization of the Dirac phenomenology optical potential~\cite{Cooper09}. It reproduces the energy and angular dependence~\cite{Matsumura26} more accurately than those used in previous studies~\cite{Yoshida16,Yoshida19,Taniguchi_48Ti,Yoshida22_Po}, and ensures a consistent treatment of the $p$-$\alpha$ cross section and the distorted waves.

The reduced transition matrix element is a convolution of the distorted waves and the $\alpha$ preformation amplitude,
\begin{align}
  \label{eq:T-matrix}
  \bar{T}   = & \int d^3 R\,
  \chi_{p     ,\bm{K}_p     }^{(-)*}(\bm{R})
  \chi_{\alpha,\bm{K}_\alpha}^{(-)*}(\bm{R})
  \chi_{p_0   ,\bm{K}_{p_0} }^{(+) }(\bm{R}) \nonumber                  \\
              & \times e^{-i\left(4/12\right)\bm{K}_{p_0} \cdot \bm{R}}
  \mathcal{Y}_\alpha(R) Y_{00}(\hat{\bm{R}}),
\end{align}
where $\chi^{(\pm)}_{i,\bm{K}_i}$ denotes the distorted wave with asymptotic wave number $\bm{K}_i$ for particle $i = p_0$, $p$, and $\alpha$, corresponding to the incident proton, scattered proton, and knocked-out $\alpha$ particle, respectively.
$\mathcal{Y}_\alpha(R)$ is the $\alpha$ preformation amplitude in the target nucleus.
Accordingly, the transition matrix element, and hence the TDX, is sensitive to the magnitude and spatial distribution of $\mathcal{Y}_\alpha(R)$.

For the $p$-$^{12}\mathrm{C}$ and $p$-$^{8}\mathrm{Be}$ distorted waves, we employ the EDAD1 parametrization of the Dirac phenomenology optical potential~\cite{Hama90,Cooper93,Cooper09}, while for the $\alpha$-$^{8}\mathrm{Be}$ channel we use the Avrigeanu parametrization~\cite{Avrigeanu94}.
To estimate the optical potential uncertainties, parameter set~I of Ref.~\cite{Mabiala09} is also considered.

In contrast to the previous phenomenological analysis, we determine the
$\alpha$ preformation amplitude directly from microscopic wave functions.
\begin{align}
  \mathcal{Y}_\alpha(R) = \sqrt{\frac{12!}{4!\,8!}}
  \Braket{
  \frac{\delta(r - R)}{R^2}\,\Phi_{\alpha}\,\Phi_{{}^{8}\mathrm{Be}}\,Y_{00}(\hat{r})
  |\Phi_{^{12}\mathrm{C}}},\label{eq:rwa}
\end{align}
where $\Phi_{\alpha}$, $\Phi_{^{8}\mathrm{Be}}$, and $\Phi_{^{12}\mathrm{C}}$
are the corresponding ground-state wave functions, and
$r$ is the relative coordinate between the $\alpha$ and $^{8}\mathrm{Be}$.
By definition, $R^2 |\mathcal{Y}_\alpha(R)|^2\, dR$ gives the probability of finding the $\alpha$-$^{8}\mathrm{Be}$ configuration at relative distance $R$.

To diagnose how the structure of $^{12}\mathrm{C}$ affects the
$\alpha$ preformation, we introduce two extreme
$^{12}\mathrm{C}$ models: $\alpha$-cluster and harmonic-oscillator
(HO)-based models.
To isolate the effect of the $^{12}\mathrm{C}$ structure, we fix the
$\alpha$ and $^{8}\mathrm{Be}$ wave functions throughout to those
obtained from the cluster model described below, whose parameters
reproduce the charge radius of the $\alpha$ particle, the
$\alpha$-$\alpha$ scattering phase shifts, and the narrow $0^+$
resonance of $^{8}\mathrm{Be}$.

  {\it $\alpha$-cluster models: }
The $\alpha$-cluster models describe nuclei as systems of $\alpha$ clusters, each represented by the following localized $(0s)^4$ Gaussian wave packet centered at the generator coordinate $\bm s$~\cite{Brink1966},
\begin{align}
  \Phi_{\alpha}(\bm{s})
                    & := \mathcal{A} \left\{ \phi(\bm{r}_1, \bm{s}) \chi_{p\uparrow} \cdots \phi(\bm{r}_4, \bm{s}) \chi_{n\downarrow} \right\},\label{eq:alpha} \\
  \phi(\bm r,\bm s) & := \left({2\nu}/{\pi}\right)^{3/4}
  \exp\left\{-\nu\left(\bm r - \bm s\right)^2 \right\},
\end{align}
where the Gaussian width parameter is set to $\hbar\omega=2\hbar^2\nu/m=22.8$ MeV,
which reproduces the charge radius of the $\alpha$ particle.

The wave function of $N\alpha$ system is described by a superposition of  various cluster
configurations. The ground state of $^{8}\mathrm{Be}$, a narrow resonance in the $2\alpha$
continuum, is described by a $2\alpha$ cluster model,
\begin{align}
  \Phi_{^{8}\rm Be} =\sum_ic_iP^{0^+}\!\!\mathcal{A} \left\{ \Phi_{\alpha}(-\bm s_i/2)\, \Phi_{\alpha}(\bm s_i/2) \right\},\label{eq:be8}
\end{align}
where $P^{0^+}$ is the projection operator onto the spin-parity $0^+$. The generator coordinate $\bm s_i=(0,0,s_i)$ is discretized from 1.0 to 15 fm with an interval of 0.5 fm.
The coefficients $c_i$ are determined by diagonalizing the Hamiltonian composed of the nucleon kinetic energies, the Coulomb interaction, and the Volkov No.~2 nucleon-nucleon interaction~\cite{Volkov1965}.
The Majorana exchange parameter is set to $M=0.595$, which reproduces
the $s$- and $d$-wave $\alpha$-$\alpha$ scattering phase shifts and the narrow
$0^+$ resonance, providing a realistic description of $^{8}\mathrm{Be}$.
Owing to its extremely small width, the $0^+$ resonance is well approximated
by a bound-state-like state over the spatial region relevant to the
present reaction~\cite{Pichler1997,Cuong2020,Descouvemont2021}.

For $^{12}\mathrm{C}$, the $\alpha$-cluster wave function is given as
\begin{align}
  \Phi_{^{12}{\rm C}}= \sum_i c_iP^{0^+}\!\!
  \mathcal{A} \left\{ \Phi_{\alpha}(\bm s^{(i)}_1)\Phi_{\alpha}(\bm s^{(i)}_2)\Phi_{\alpha}(\bm s^{(i)}_3) \right\},\label{eq:3a}
\end{align}
where the generator coordinates $s^{(i)} := \{\bm s_1^{(i)},\bm s_2^{(i)},\bm s_3^{(i)}\}$ are generated by the time-dependent cluster model and sufficiently sampled to achieve convergence~\cite{Imai19}.
The Hamiltonian is identical to that for $^{8}{\rm Be}$, except for the Majorana exchange parameter, which is set to $M=0.587$ to simultaneously reproduce the charge radius (2.47 fm) and $\alpha$ separation energy (7.27 MeV). In the following, this model is referred to as the unrestricted $3\alpha$ model.

We also consider an equilateral-triangle version of $\alpha$-cluster model as in the algebraic cluster model~\cite{Bijker2020}, in which the wave function is still given by Eq.~\eqref{eq:3a}, but the generator coordinates $\bm s^{(i)}$ are restricted to equilateral-triangular configurations.
The side length of the triangle is discretized from 0.5 to 15 fm with an interval of 0.5 fm.
The Majorana parameter is set to $M = \mathrm{0.587}$ to reproduce the $\alpha$ separation energy.
In the following, this model is referred to as the equilateral $3\alpha$ model.\\
\begin{figure}[h]
  \centering
  \includegraphics[width=\hsize]{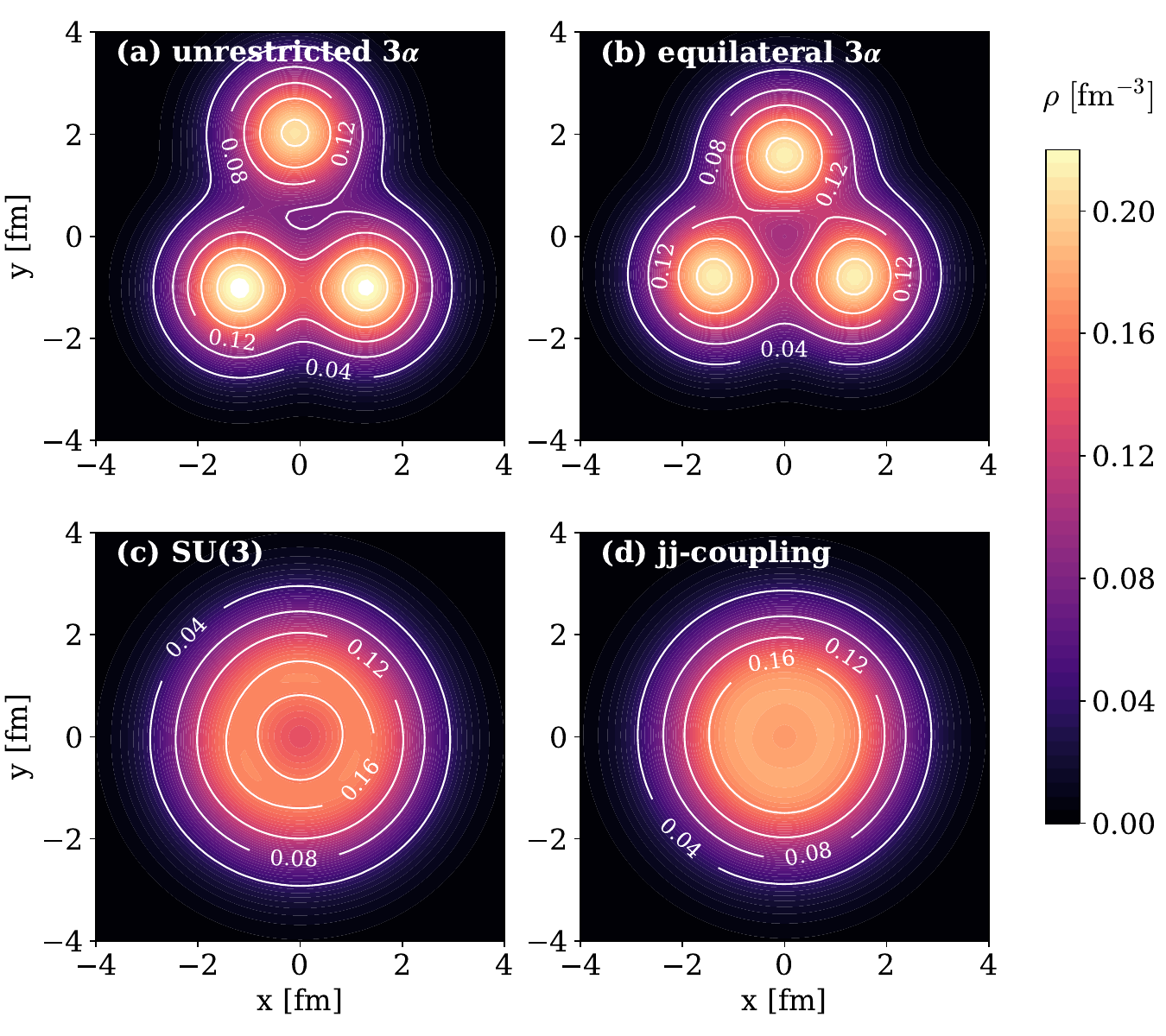}
  \caption{Density distributions of the representative configurations for (a) unrestricted $3\alpha$, (b) equilateral $3\alpha$, (c) SU(3), and (d) $jj$-coupling models.
    For the unrestricted and equilateral $3\alpha$ models, the representative configurations are chosen as those corresponding to the generator coordinates that have the largest overlap with the GCM wave function [Eq.~\eqref{eq:3a}].}
  \label{fig:density}
\end{figure}

{\it HO-based configurations: }
As a limit without explicit cluster correlation, we introduce two HO-based configurations.
The first one is the SU(3) limit where the spin-orbit splitting is neglected
and the ground state is described by the
$(0s)^4(0p_x)^4(0p_y)^4$ configuration with an oblate-deformed shape.
The oscillator parameter is chosen as $\hbar\omega=15.59$ MeV to reproduce
the charge radius of $^{12}\mathrm{C}$.
The second is the $jj$-coupling limit which introduces the spin-orbit splitting, where the ground-state is represented by a $(0s)^4(0p_{3/2})^8$ configuration.
This configuration is commonly regarded as the zeroth-order shell-model description of $^{12}\mathrm{C}$.
The oscillator parameters are chosen anisotropically to  reproduce the charge radius of the ground state and the quadrupole moment of the $2^+$ state simultaneously;
$\hbar\omega_z = 10.8\ \mathrm{MeV}$ and
$\hbar\omega_x = \hbar\omega_y = 32.6\ \mathrm{MeV}.$

Figure~\ref{fig:density} shows the density distributions of the representative configurations.
The cluster and HO-based models exhibit qualitatively different density distributions: the former show localized density reflecting developed $\alpha$ clustering, whereas the latter exhibit smooth distributions without clear cluster localization.
Within the cluster models, the unrestricted $3\alpha$ model shows a slight breaking of equilateral-triangular symmetry associated with the $^{8}\mathrm{Be}$-$\alpha$ correlation.

\begin{figure}[h]
  \includegraphics[width=\hsize]{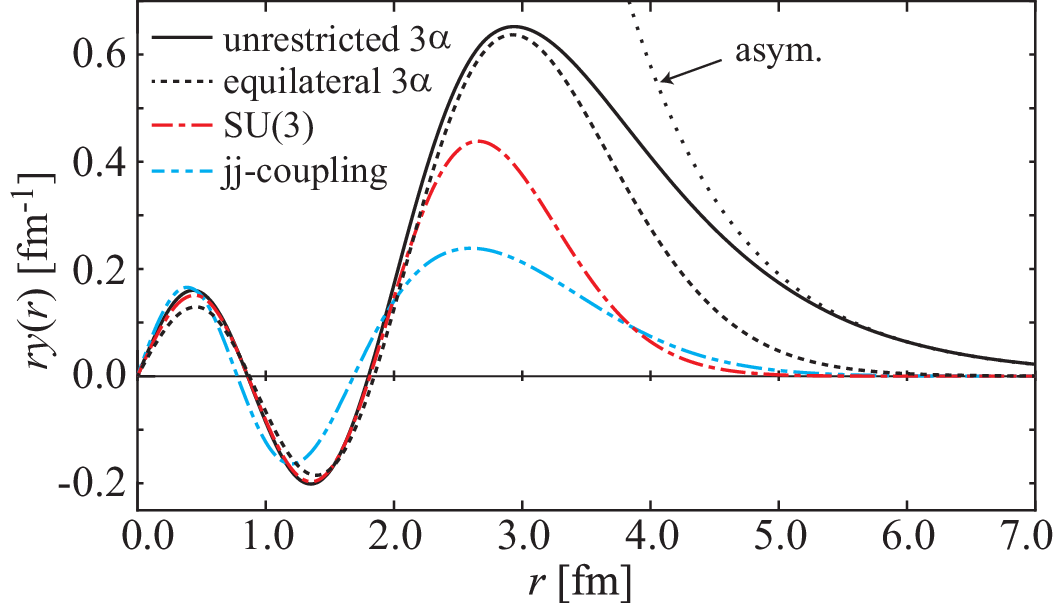}
  \caption{\label{fig:rwa}
    The $\alpha$ preformation amplitudes obtained from $\alpha$-cluster and HO-based models. The dotted line shows the asymptotic behavior described by the Whittaker function.}
\end{figure}
Figure~\ref{fig:rwa} shows the $\alpha$ preformation amplitudes evaluated with Eq.~\eqref{eq:rwa}, where the $\alpha$ and $^{8}\mathrm{Be}$ wave functions are given by Eqs.~\eqref{eq:alpha} and \eqref{eq:be8}, respectively, while the four models introduced above are employed for $^{12}\mathrm{C}$.
As expected, the $\alpha$-cluster models yield large and spatially extended $\alpha$ preformation amplitudes, whereas those obtained with the HO-based descriptions are much smaller and shorter-ranged.
Within the HO-based descriptions, the $jj$-coupling model yields the smallest amplitude, because the nucleon spin is not saturated and the overlap with the $\alpha+\rm {}^{8}Be$ channel is reduced.

The asymptotic behavior also shows strong model dependence.
At large distances where the nuclear interaction and antisymmetrization are negligible, it should approach the Whittaker function. The unrestricted 3$\alpha$ model naturally reproduces this behavior, whereas the equilateral $3\alpha$ model does not because the equilateral-triangular constraint prevents the $\alpha+{}^{8}\mathrm{Be}$ asymptotic configuration.
In the HO-based models, the tail is governed by the Gaussian form of the HO basis, resulting in rapid damping.
We also note that the cluster models exhibit similar $\alpha$ preformation amplitudes in the interior region ($r\lesssim3~\mathrm{fm}$), indicating that the $3\alpha$ system naturally favors an equilateral-triangular configuration.

\begin{figure*}
  \includegraphics[width=\hsize]{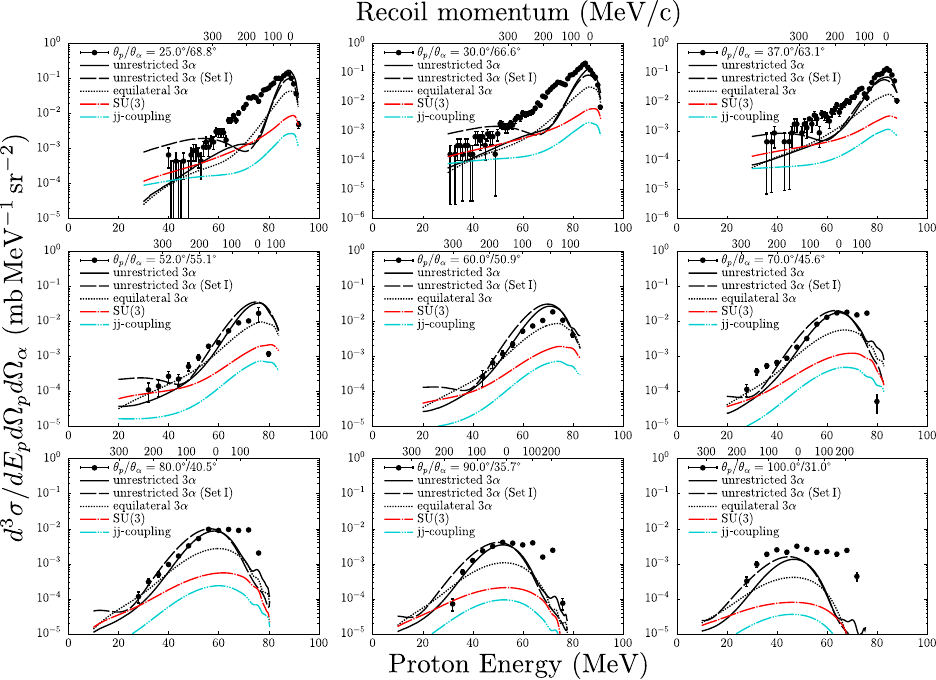}
  \caption{\label{fig:TDX_all}
    Comparison between the calculated TDXs and the experimental data~\cite{Mabiala09}.
    Each panel corresponds to a different kinematical condition specified by
    the emission angles of the proton and $\alpha$ particle, $\theta_p/\theta_\alpha$.
    The lower and upper horizontal axes show the emitted proton energy $T_p$ and the recoil momentum $Q$, respectively.
    The result labeled ``set~I'' is obtained with the unrestricted
    $3\alpha$ model using optical-potential set~I of Ref.~\cite{Mabiala09}.
  }
\end{figure*}

The $^{12}\mathrm{C}(p,p\alpha)^{8}\mathrm{Be}$ TDXs are calculated with
the DWIA code {\sc pikoe}~\cite{Ogata23} and compared in Fig.~\ref{fig:TDX_all}
with the experiment~\cite{Mabiala09}.
Each panel corresponds to a different kinematics specified by
the proton and $\alpha$ emission angles, $\theta_p/\theta_\alpha$.
Unless otherwise stated, the EDAD1~\cite{Hama90,Cooper93,Cooper09} and Avrigeanu parametrizations~\cite{Avrigeanu94} are
used for the proton and $\alpha$ distorted waves, respectively.

We first focus on the cross-section peaks near $Q\simeq0$.
The importance of the peak height can be understood qualitatively from the
plane-wave limit, where the reduced transition matrix becomes the Fourier
transform of the $\alpha$ preformation amplitude~\cite{Wakasa17},
\begin{align}
  \bar{T}^{\mathrm{PW}}(Q) =
  \int d^3R\, e^{i\bm{Q}\cdot\bm{R}}
  \mathcal{Y}_{\alpha}(R)Y_{00}(\hat{\bm{R}})
\end{align}
Here, $\bm{Q}=(1-4/12)\bm{K}_{p_0}-\bm{K}_p-\bm{K}_\alpha$
is the recoil momentum of the residual nucleus $^{8}\mathrm{Be}$.
At the recoilless condition $Q=0$, this reduces to the radial
integral
$  \bar{T}^{\mathrm{PW}}(0)
  = \sqrt{4\pi}\int_0^\infty dR\,R^2\mathcal{Y}_\alpha(R).$
In the nuclear interior, the preformation amplitude oscillates due to the
Pauli principle, and these contributions largely cancel in the integral,
leaving the outer part dominant.
In the DWIA calculation, the distorted waves further suppress contributions
from the interior and enhance the peripheral sensitivity.
Consequently, the cross-section peak primarily probes the $\alpha$
preformation in the nuclear surface.

Keeping this in mind, we focus on the middle and lower panels of
Fig.~\ref{fig:TDX_all}, for which the DWIA description of quasi-free
knockout is expected to be more reliable owing to the large momentum
transfer to the $\alpha$ particle.
In these panels, the unrestricted $3\alpha$ model clearly gives the
closest description of the data, reproducing the peak heights
and the local $T_p$ and $Q$ dependence around the peaks.
By contrast, the HO-based models underestimate the data by more than an
order of magnitude.
Thus, unlike Mabiala et al.~\cite{Mabiala09}, we conclude that the data support a nearly fully developed three-$\alpha$-cluster structure represented by the unrestricted $3\alpha$ model, and reject the conventional HO-based descriptions.

The reason for this different conclusion is as follows: Mabiala et al. extracted an $S$ factor from the measured cross sections and found it consistent with a $p$-shell estimate~\cite{Kurath1973}, but did not compare the spatial distribution of the preformation amplitude. However, as shown here, a sufficiently large preformation amplitude in the nuclear surface region is essential for reproducing the cross sections. With the realistic $^{8}\mathrm{Be}$ wave function, the HO-based $p$-shell descriptions yield amplitudes that are too small and short-ranged to reproduce the data.

Although the unrestricted $3\alpha$ model describes the peak region well,
notable discrepancies remain in the upper panels and on the high-$T_p$ side of the lower panels of Fig.~\ref{fig:TDX_all}.
Part of these discrepancies may originate from the contribution of the broad $^{8}\mathrm{Be}(2^+)$ resonance within the experimental gate. Including the estimated $2^+$ contribution improves the unrestricted $3\alpha$ model descriptions, whereas the SU(3) prediction remains far below
the data (see Supplemental Material).
The remaining discrepancies likely reflect the limitations of the one-step DWIA description.
In the lower panels, the calculation tends to underestimate the data on the high-$T_p$ side, where the emitted $\alpha$ energy is low, and final-state interactions between the $\alpha$ particle and the residual nucleus may become important.
Such effects, including possible channel coupling, are beyond the DWIA framework.
A similar discrepancy at high $T_p$ has been reported in $^{16}\mathrm{O}(p,p\alpha)$ and $^{40}\mathrm{Ca}(p,p\alpha)$ analyses at around 100~MeV~\cite{Carey84}.
By contrast, recent $^{40}\mathrm{Ca}(p,p\alpha)$ data at
392~MeV~\cite{Matsumura26} show improved agreement with DWIA calculations,
probably because the higher incident energy makes the one-step quasi-free
knockout picture more appropriate.
This suggests that measurements at higher energies and over a wider recoil-momentum range would be valuable.

The upper panels exhibit a similar limitation.
In these forward-angle kinematics, the calculated TDXs have a dip around
$Q\sim200~\mathrm{MeV}/c$ due to the node structure of $\bar{T}^{\mathrm{PW}}(Q)$.
However, the experimental data do not show such a pronounced dip and decrease more gradually. This discrepancy is also likely caused by the kinematical conditions of the upper panels. For a one-step DWIA description, small recoil momentum $Q$ and moderate proton emission energy, or equivalently sufficiently high $\alpha$ emission energy, are desirable. These conditions are not simultaneously satisfied in the upper panels having forward $p$-$\alpha$ scattering kinematics. Indeed, the same difficulty was seen in the original analysis of Ref.~\cite{Mabiala09}, where these angular settings were not reproduced satisfactorily.

Finally, we examine the uncertainty arising from the optical potentials.
For this purpose, we repeat the calculation using the same $\alpha$ preformation amplitude
from the unrestricted $3\alpha$ model, but replacing the optical potentials by parameter set~I
of Ref.~\cite{Mabiala09}.
The optical-potential dependence is most visible in the upper panels, especially on the low-$T_p$ side, and may partly account for the discrepancy in this region.
The peak heights at $Q\simeq 0$ also change slightly in almost all panels.
However, these changes are much smaller than the differences caused by the $\alpha$ preformation amplitudes.
Thus, the conclusion that the experimental data require the developed $3\alpha$-cluster limit is not affected by this uncertainty.

In summary, the ground state of $^{12}\mathrm{C}$ has long been approximated by a mean-field picture, although the presence of $\alpha$-cluster correlations remains an open question due to the limited discriminating power of conventional observables.
In this work, we examine the $^{12}\mathrm{C}(p,p\alpha)^{8}\mathrm{Be}$ reaction as a direct probe of
$\alpha$-cluster formation by combining DWIA calculations with $\alpha$ preformation amplitudes derived from three-$\alpha$-cluster models and HO-based descriptions.
The calculated cross sections show strong model dependence: the unrestricted 3$\alpha$-cluster model reproduces the experimental data, whereas the HO-based models underestimate the data by more than an order of magnitude.
Thus, although the previously extracted spectroscopic factor was consistent
with a shell-model estimate, a direct microscopic calculation of the
preformation amplitude shows that the same data require a spatially
developed $3\alpha$ configuration.

\textit{Acknowledgments}-- We wish to acknowledge S.~Ogawa, Y.~Chazono and K.~Ogata for valuable comments and discussions. This work is supported in part by Grant-in-Aid for Scientific Research (No.~JP25K17400 and JP26K00703) from Japan Society for the Promotion of Science (JSPS), and JST ERATO Grant No. JPMJER2304, Japan.



\bibliography{apssamp}

\end{document}